\documentclass[prl,twocolumn,floats,aps,epsfig,nofootinbib,amssymb]{revtex4}
\usepackage{subfigure}
\usepackage{graphicx}
\usepackage{cancel}
\usepackage{amssymb}
\usepackage{textcomp}
\usepackage{amsmath}
\usepackage{bm}
\usepackage{times}
\usepackage{epsfig}
\usepackage{color}
\def\beq{\begin{equation}}
\def\eeq{\end{equation}}
\def\bea{\begin{eqnarray}}
\def\eea{\end{eqnarray}}

	\newcommand{\abs}[1]{ \mathopen{}\left| {#1}\right| }
	\DeclareMathOperator{\real}{Re}
	\DeclareMathOperator{\imag}{Im}
	
\begin{document}

\title{\large Spontaneous R-Parity Breaking in SUSY Models}
\bigskip
\author{Pavel Fileviez P{\'e}rez}
\author{Sogee Spinner}
\address{
Department of Physics, University of Wisconsin, Madison, WI 53706, USA}
\date{\today}

\begin{abstract}
We investigate a mechanism for spontaneous R-parity breaking in a
class of extensions of the minimal supersymmetric standard model with an
extra Abelian gauge symmetry which is a linear combination of $B-L$ and
weak hypercharge. Both $U(1)_{X}$ and R-parity are broken by the vacuum
expectation value of the right-handed sneutrinos which is proportional
to the soft SUSY masses. In these models the mechanism
for spontaneous R-parity violation can be realized even with positive
soft masses. In this context one has a realistic mechanism for generating
neutrino masses as well as a realistic spectrum. We briefly discuss the
possible collider signals which could be used to test the theory, the
contributions to proton decay and the possibility of a
gravitino as a dark matter candidate.
\end{abstract}

\maketitle

\section{I. Introduction}
There are several ideas in physics beyond the standard model 
for protecting the Higgs mass but Supersymmetry (SUSY) is perhaps
the most appealing. It is well-known that in order to make
predictions in the context of the minimal supersymmetric
Standard Model (MSSM) one has to understand the origin
of SUSY breaking and the R-parity violating terms.
For a review on Supersymmetry see Ref.~\cite{Haber}.

The so-called R-parity discrete symmetry is defined as
$R=(-1)^{ 3(B-L) + 2S}=(-1)^{2S}M$, where the $B$, $L$, and
$S$ stand for the baryon number, lepton number and
spin, respectively. Here $M$ is the so-called matter parity
which is $-1$ for any matter superfield and $+1$ for any
Higgs or Gauge superfield. See Ref.~\cite{Martin} for
the studies of R-parity conservation in SUSY models at
the low-scale and Ref.~\cite{Goran} in the context
of $SO(10)$ models. For studies of the phenomenological 
aspects of SUSY models with R-parity violation 
see Ref.~\cite{Tao}.

In order to understand the origin of the R-parity violating 
operators present in the MSSM one has two
possibilities: i) one can look for an extension
of the MSSM where $B-L$ is a global symmetry~\cite{GlobalBL},
but one has to face the Majoron problem~\cite{Majoron}.
ii) study this issue in a framework where $B-L$ is a
local symmetry. In this case the Majoron is eaten by the
new gauge bosons in the theory and a simple
consistent framework exists.

Recently, this latter option was studied in the context
of left-right symmetric models~\cite{FileviezPerez:2008sx} and
in a simple extension of the MSSM where one has an extra $B-L$
Abelian gauge symmetry~\cite{Barger:2008wn}. In this paper we
want to study this approach to spontaneous R-parity breaking in
the simplest most general extension of the MSSM where the
new Abelian symmetry is a linear combination of the weak
hypercharge and $B-L$. In this way all
anomalies are canceled by adding only right-handed neutrinos.
We discuss the full spectrum of the theory and the possible signals
at the LHC. In this context the mechanism
for spontaneous R-parity violation is possible even with
positive soft-terms, in contrast with the results in
Ref.~\cite{FileviezPerez:2008sx} and Ref.~\cite{Barger:2008wn}
where one needs negative mass soft terms for the ``right-handed"
sneutrinos.

This paper is organized as follows: In Section II we discuss
the mechanism for Spontaneous R-parity Breaking in a simple
extension of the MSSM. In Section III we discuss the R-parity
violating terms obtained and the full spectrum of the theory,
while in Section IV the possible collider signals are pointed
out. In Section V we discuss the possibility of
gravitino cold dark matter and the proton decay
constraints. Finally, we summarize our findings in Section VI.
\section{II. Spontaneous R-parity Breaking}
The well-known R-parity violating terms in the MSSM are
\begin{eqnarray}
\label{W.RPV}
{\cal W}_{RpV} &=& \epsilon_i \hat{L}_i \hat{H}_u \ + \  \lambda_{ijk} \hat{L}_i \hat{L}_j \hat{E}^C_k
\nonumber \\
& + & \lambda_{ijk}^{'} \hat{Q}_i \hat{L}_j \hat{D}^C_k 
 \ + \ \lambda_{ijk}^{''} \hat{U}^C_i \hat{D}^C_j \hat{D}^C_k 
\end{eqnarray}
where $\lambda_{ijk}^{''}=-\lambda_{ikj}^{''}$ and $\lambda_{ijk}=-\lambda_{jik}$.
In the above equation the first three interactions violate 
lepton number, while the last term violates baryon number.

The main goal of this paper is the investigation of the mechanism
for spontaneous R-parity violation in simple local Abelian extensions of
the MSSM, which contain $B-L$. Then, in this context the new hypercharge will
be defined as
\begin{equation}
X = a \ Y \ + \ b \ \left( B \ - \ L \right),
\end{equation}
where $Y$, $B$ and $L$ stand for the weak hypercharge,
baryon number and lepton number, respectively. We stick to
this simple linear combination since one can easily show
that only three right-handed nuetrinos are necessary for anomaly cancelation.

\subsection{II. A. Anomaly Cancellation}
\label{Sec.Anomaly.Cancellation}
As it is well-known once we extend the SM gauge group one
introduces constraints on the particle charges based
on anomaly cancelation. For the MSSM particle content
with an $n_N$ additional right-handed neutrinos,
the $\left( SU(2)_L, U(1)_Y, U(1)_X \right)$ charge
assignment is:
\begin{equation*}
\hat{Q} = \left(
\begin{array} {c}
\hat{U} \\ \hat{D}
\end{array}
\right) \ \sim \ (2, \frac{1}{3}, X_Q),
\ \ \
\hat{L} = \left(
\begin{array} {c}
 \hat{N} \\ \hat{E}
\end{array}
\right) \ \sim \ (2, -1, X_L),
\end{equation*}
\begin{equation*}
\hat{H}_u = \left(
\begin{array} {c}
	\hat{H}_u^+
\\
	\hat{H}_u^0
\end{array}
\right) \ \sim \ (2, 1, X_H),
\end{equation*}
\begin{equation*}
\hat{H}_d = \left(
\begin{array} {c}
	\hat{H}_d^0
\\
	\hat{H}_d^-
\end{array}
\right) \ \sim \ (2, -1, - X_H),
\end{equation*}
\begin{equation*}
\hat{U}^C \ \sim \ (1,-\frac{4}{3}, X_{U}),
\ \ \
\hat{D}^C \ \sim \ (1,\frac{2}{3}, X_{D}),
\ \ \
\hat{E}^C \ \sim \ (1,2, X_{E}),
\end{equation*}
and
\begin{equation*}
\hat{N}^C \ \sim \ (1,0, X_{N}).
\end{equation*}
In this case the following anomaly
constraints must be satisfied:
\begin{align}
	\label{X.Color}
	U(1)_X \left[SU(3)_C\right]^2 : & \quad 2 X_Q + X_{U} + X_{D} = 0,
	\\
	\label{X.Isospin}
	U(1)_X \left[SU(2)_L\right]^2 : & \quad 3 X_Q + X_L = 0,
	\\ \notag
	U(1)_X \left[U(1)_Y\right]^2 : & \quad 3 \left(2 Y_Q^2 X_Q + Y_{U}^2 X_{U} + Y_{D}^2 X_{D} \right)
	\\
	\label{X.Y2}
	& \quad + 2 Y_L^2 X_L + Y_{E}^2 X_{E}= 0,
	\\ \notag
	U(1)_X^2 \left[U(1)_Y\right] : & \quad 3 \left(2 Y_Q X_Q^2 + Y_{U} X_{U}^2 + Y_{D} X_{D}^2 \right)
	\\
	\label{X2.Y}
	& \quad + 2 Y_L X_L^2 + Y_{E} X_{E}^2 = 0,
	\\ \notag
	U(1)_X^3 : & \quad 3 N_g \left(2 X_Q^3 + X_{U}^3 + X_{D}^3 \right)  + n_{N} X_{N}^3
	\\
	\label{X3}
	& \quad + N_g \left(2 X_L^3 + X_{E}^3\right)= 0,
	\\ \notag
	U(1)_X : & \quad 3 N_g \left(2 X_Q + X_{U} + X_{D} \right)  + n_{N} X_{N}
	\\
	\label{X}
	& \quad + N_g \left(2 X_L + X_{E}\right)  = 0,
\end{align}
where the hypercharge of a field $\phi$ is given by $Y_\phi$.
The first five equations are mixed gauge anomalies, while the last one is a
gauge-gravity anomaly. These are similar to equations derived before, for example
in~\cite{Chamseddine:1995rs, EmmanuelCosta:2009za}.  In the above conditions
$N_g$ is the number of generations and $N_g = 3$ will be assumed
for the rest of this Letter. Higgs charges cancel in the above automatically
because of their opposite charges. It is straightforward to show that if the
new Abelian symmetry is a linear combination of the weak hypercharge and
$B-L$ all anomalies cancel.

Using the equations linear in the $X$-charges (Eqs. (\ref{X.Color}) - (\ref{X.Y2}) and (\ref{X}))
it is possible to express $X_Q, \ X_L, \ X_{U} \ \text{and} \ X_{D}$ in terms of $X_{E}$ and $X_{N}$:
\begin{align}
	\label{xQ}
	X_Q & = \frac{1}{18} \left(3 X_E + n_N X_N \right),
\\
	\label{xL}
	X_L & = - \frac{1}{6} \left(3 X_E + n_N X_N \right),
\\
	\label{xU}
	X_U & = \frac{1}{9} \left(- 6 X_E + n_N X_N \right),
\\
	\label{xD}
	X_D & = \frac{1}{9} \left(3 X_E - 2 n_N X_N \right).
\end{align}
Substituting these into Eq. (\ref{X3}) yields:
\begin{equation}
	n_N - \frac{1}{9}n_N^3 = 0.
\end{equation}
Therefore, $n_N = 3$, and this value will be used from now on. A similar substitution
into Eq. (\ref{X2.Y}) reveals no new constraints.

Further constraints are contributed from the couplings in the superpotential.
Since
\begin{equation}
	\label{W}
	{\cal W} = {\cal W}_{MSSM} \ + \ Y_\nu^D \ \hat{L}^T \ i \sigma_2 \ \hat{H}_u \ \hat{N}^C,
\end{equation}
where
\begin{eqnarray}
	{\cal W}_{MSSM} &=& Y_u \ \hat{Q}^T \ i \sigma_2 \ \hat{H}_u \hat{U}^C
\ + \ Y_d \ \hat{Q}^T \ i \sigma_2 \ \hat{H}_d \hat{D}^C \nonumber \\
& + & Y_e \ \hat{L}^T \ i \sigma_2 \ \hat{H}_d \ \hat{E}^C
\ + \ \mu \ \hat{H}_u^T \ i \sigma_2 \ \hat{H}_d,
\end{eqnarray}
the Higgs $X$-charge is:
\begin{equation}
	\label{xH}
	X_H = \frac{1}{2} \left(X_E - X_N\right).
\end{equation}

All superpotential interactions are consistent with Eqs. (\ref{xQ}) - (\ref{xD}) and (\ref{xH}).
At this point it is illuminating to state all the conditions on the charges in terms of $a \equiv X_H$
and $b \equiv X_N$:
\begin{align}
	X_Q & = \frac{1}{3}a + \frac{1}{3}b,
	\\
	X_L & = -a -  b,
	\\
	X_U & = -\frac{4}{3} a - \frac{1}{3}b,
	\\
	X_D & = \frac{2}{3} a - \frac{1}{3} b,
	\\
	X_E & = 2 a + b,
\end{align}
or simply stated: $X = a \ Y \ + \ b \ (B-L)$.  Since both $Y$ and $X$ will be separately conserved, $B-L$ will also be conserved making R-parity 
an exact symmetry as well. Therefore, the most general charge assignment possible is also consistent with the goal discussed in the introduction.

Some interesting cases are: $a=0$ and $b=1$, which corresponds to $B-L$;
$a=1$ and $b=-5/4$ which is the GUT normalization for this group and
allows it to be embedded in $SO(10)$.  Note that $b=0$ is just a scaled
version of hypercharge and does not constitute a new charge. It is
possible to have different charges for the $\hat N$ \cite{Chamseddine:1995rs},
but here we are interested in the case where $B-L$ is part of the symmetry.
\subsection{II. B. Symmetry Breaking}
The particle content, charge assignment and superpotential (Eq.~(\ref{W}))
where all given in the previous section.  Symmetry breaking is achieved through the
right-handed sneutrinos, which have a non-trivial $X$-charge.
Once one of these fields acquire a vacuum expectation value (VEV), it spontaneously breaks both
the gauge symmetry, $U(1)_{X}$, as well as R-parity and forces left-handed sneutrino,
through mixing terms, to acquire a VEV. Since $B-L$ is part of the gauge
symmetry the Majoron~\cite{Majoron} (the Goldstone boson associated with spontaneous
breaking of lepton number) becomes the longitudinal component of the $Z^{'}$
and does not pose a problem, thus allowing for a general,
simple mechanism for spontaneous R-parity breaking.

In addition to the superpotential, the model is also specified
by the soft terms:
\begin{eqnarray}
	\nonumber
	V_{soft} & = & M_{\tilde N^C}^2 \abs{\tilde{N}^C}^2 \ + \ M_{\tilde L}^2 \ \abs{\tilde L}^2 + M_{\tilde E^C}^2 \ \abs{\tilde E^C}^2
	\\
	\nonumber
		& + & m_{H_u}^2 \abs{H_U}^2 + m_{H_d}^2 \abs{H_D}^2 \ + \ \left( \frac{1}{2} M_{X} \tilde{B^{'}} \tilde{B^{'}} \right.
  	\nonumber
	\\
		& + &  \left. A_\nu^D \ \tilde{L}^T \ i \sigma_2 \ H_u \ \tilde{N}^C  \  + \  B\mu \ H_U^T \ i \sigma_2 \ H_D \right.
	\nonumber \\
		& + & \left. \mathrm{h.c.} \right) +...
\label{soft}
\end{eqnarray}
where the terms not shown here correspond to terms in the soft MSSM potential.
Now, we are ready to investigate the predictions of this mechanism
for spontaneous R-parity violation.

Here, symmetry breaking is achieved through the VEVS of sneutrinos ($\langle \tilde{\nu_i} \rangle=v_L^i/\sqrt{2}$ and $\langle \tilde\nu^C_i \rangle=v_R^i/\sqrt{2}$) and the Higgs doublets
($\left< H_u^0 \right> = v_u/\sqrt{2}$ and $\left< H_d^0 \right> = v_d/\sqrt{2}$).

The scalar potential in this theory is given by
\begin{eqnarray}
	V & = & V_{F} \ + \ V_D \ + \ V_{soft}^S,
\end{eqnarray}
where the relevant terms for $V_{soft}^S$ are given
in Eq.~(\ref{soft}). Once one generation of sneutrinos,
$\tilde{\nu}$ and $\tilde{\nu}^C$, and the Higgses, acquire
a VEV, the potential reads
\begin{align}
	\left<V_F \right> = &
		\frac{1}{4} \left(Y_\nu^D \right)^2
		\left(
			v_R^2 v_u^2 + v_R^2 v_L^2 + v_L^2 v_u^2
		\right)
		\nonumber \\
		&+ \frac{1}{2} \mu^2
		\left(
			v_u^2 + v_d^2
		\right)
		 -  \frac{1}{\sqrt{2}} Y_\nu^D \ \mu \ v_R v_L v_d,
	\\
	\left<V_D \right> =&
		\frac{1}{32}
		\left[
			g_2^2
			\left(
				 v_u^2 -v_d^2 - v_L^2
			\right)^2
			+ g_1^2
			\left(
				v_u^2 - v_d^2 - v_L^2
			\right)^2
		\right.
		\nonumber \\
		& +  \left.
			g_{X}^2
			\left(
				b v_R^2 - (a+b) v_L^2 + a v_u^2 - a v_d^2
			\right)^2
		\right],
	\\
	\left<V_{soft}^S \right> =&
		\frac{1}{2} M_{\tilde L}^2 v_L^2 + \frac{1}{2} M_{\tilde N^c}^2 v_R^2 + \frac{1}{2} M_{H_u}^2 v_u^2 + \frac{1}{2} M_{H_d}^2 v_d^2
		\nonumber \\
		& + \frac{1}{2 \sqrt{2}} \left(A_\nu^D + \left(A_\nu^D \right)^\dagger \right) \ v_R v_L v_u
		\nonumber \\
		&-  \text{Re} \left( B \mu \right) \ v_u v_d,
\end{align}
where $g_1$, $g_2$ and $g_{X}$ are the gauge couplings
for $SU(2)_L$, $U(1)_Y$ and $U(1)_{X}$ respectively. Minimizing
in the limit $v_R, v_u, v_d \gg v_L$:
\begin{align}
	\label{MC.vR}
	v_R & =
		\sqrt{\frac{- 8 M_{\tilde{N}^c}^2 - a b \ g_X^2 \left(v_u^2 - v_d^2\right)}{g_{X}^2 b^2}},
	\\
	v_L & =
		\frac{B_\nu v_R}{M_{\tilde L}^2 - \frac{1}{8} g_{X}^2 \left(a+b\right) b \ v_R^2 - D_{\text{ew}}^2
		},
	\\
	|\mu|^2 & = -\frac{1}{2} m_Z^2 - \frac{1}{8} a^2 g_X^2 \left(v_u^2 + v_d^2\right) - \frac{M_{H_u}^2 \tan^2 \beta - M_{H_d}^2}{\tan^2 \beta-1},
	\\
	\label{MC.b}
	b & = \frac{\sin 2\beta}{2}\left(2 \mu^2 + m_{H_u}^2 + m_{H_d}^2\right),
\end{align}
where we make use of the following definitions:
\begin{align}
	B_\nu & \equiv \frac{1}{\sqrt{2}} \left(Y_\nu^D \ \mu \ v_d \ - \ A_\nu^D \ v_u \right),
	\\
	D_{\text{ew}}^2 & \equiv \frac{1}{8}\left(g_1^2 + g_2^2 + a \left(a+b\right)g_X^2\right)\left(v_u^2 - v_d^2\right),
	\\
	M_{H_u}^2 & \equiv m_{H_u}^2 + \frac{1}{8} a \ b \ g_X^2 \ v_R^2,
	\\
	M_{H_d}^2 & \equiv m_{H_u}^2 - \frac{1}{8} a \ b \ g_X^2 \ v_R^2.
\end{align}
Now, Eq.~(\ref{MC.vR}) indicates two possible scenarios
for spontaneous R-parity violation:

\begin{itemize}

\item $M_{{\tilde N}^C}^2 < 0$. This case has been studied
before in Ref.~\cite{FileviezPerez:2008sx} and Ref.~\cite{Barger:2008wn} 
in the context of left-right symmetric models and in the minimal 
gauged $U(1)_{B-L}$ extension of the MSSM. 

\item $M_{{\tilde N}^C}^2$ very small and $ab < 0$.
In this case one should satisfy the condition
\begin{equation}
|ab|  \ g_X^2 \ \left( v_u^2 - v_d^2 \right) \ > \ 8 \ M_{{\tilde N}^C}^2.
\end{equation}
This is possible for a very small $M_{\tilde N^C}^2$, as may arise in gauge
mediated supersymmetry breaking without $X$-charged messengers.

Using the constraint: $M_{Z^{'}}/g_X \geq 1$ TeV one arrives at the condition
\begin{equation}
\frac{1}{2} \ |ab|^{1/2} \ v \ \left( \frac{ \tan^2 \ \beta \ - \ 1}{ 1 \ + \ \tan^2 \ \beta } \right)^{1/2} \ \geq \ 1 \ \text{TeV}.
\end{equation}
Then, for large $\tan \beta$, $|b| \ \geq \ 66/ |a|$. This is the main constraint 
that we find for this class of models. Notice that we are using the same normalization for 
both $U(1)$ couplings in the theory and neglecting the mixing kinetic terms.
\end{itemize}
Therefore it is possible to realize
this mechanism for spontaneous R-parity violation even with 
positive soft mass terms for the right-handed neutrinos.
This possibility is quite appealing in our opinion.

\section{III. RpV Interactions and The Spectrum}
The effective MSSM-like theory will contain R-parity violating bilinear terms.
For example, the $Y_\nu^D \ \hat{L}^T \ i \sigma_2 \ \hat{H}_u \ \hat{N}^C$
term in the superpotential, leads to $Y_\nu^D \ l \ \tilde{H}_u \ v_R / \sqrt{2}$ (the $\epsilon_i$ term in Eq.~(\ref{W.RPV})) and
$Y_\nu^D  \ \tilde{H}_u^0  \ \nu^C \ v_L/ \sqrt{2}$.  The kinetic term of the lepton doublet
produces mixing between the neutrinos and neutral gauginos:
$\tilde{W}_3^0 \ \nu \ v_L / \sqrt{2}$, $\tilde{B} \ \nu \ v_L/\sqrt{2}$ and
$\tilde{B}^{'} \ \nu \ v_L / \sqrt{2}$.  While the kinetic term for right-handed neutrinos
contains the term $\tilde{B}^{'} \ \nu^C \ v_R / \sqrt{2}$. Other R-parity violating interactions
between the charged leptons and the charged components of the gauginos and Higgsinos
can be found in a similar fashion.  It is important to emphasize that all the R-parity violating
terms will be defined by two VEVs: $v_L$ and $v_R$, where $v_R \ \gg \ v_L$.
Notice that in this context one generates only bi-linear R-parity violating
terms which violate lepton number.
\subsection{III. A. Mass Spectrum}
{\subsubsection{\bf III.A.1. Gauge Bosons}}
The gauge sector consists of the SM gauge bosons and an extra neutral gauge boson,
the $Z^{'}$.  In the gauge basis $\left(B, W^0, B'\right)$, the mass matrix reads as
\begin{equation}
\label{Z.Mass.Matrix}
{\cal M}_{0}^2=\begin{pmatrix}
	\frac{1}{4} g_1^2 v^2	&	-\frac{1}{4} g_1 g_2 v^2     &     \frac{1}{4} a \ g_1 g_X v^2
	\\
	-\frac{1}{4} g_1 g_2 v^2	&	\frac{1}{4}g_2^2 v^2	   &     -\frac{1}{4} a \ g_2 g_X v^2
	\\
	\frac{1}{4} a \ g_1 g_X v^2 &	-\frac{1}{4} a \ g_2 g_X v^2 &    \frac{1}{4} g_X^2 \left(b^2 v_R^2 + a^2 v^2\right)
\end{pmatrix}.
\end{equation}
Here $v^2=v_u^2 + v_d^2 = (246)^2 \text{GeV}^2$. To satisfy the experimental constraint coming from the rho-parameter, $\rho \sim 1$,
the $Z$ mass should not be significantly modified from its MSSM expression.
Therefore, aside from the zero eigenvalue corresponding to the photon,
the eigenvalues of the above matrix are:
\begin{align}
	m_Z^2 & = \frac{1}{4}\left(g_1^2 + g_2^2\right) v^2 \left(1 - \epsilon^2\right) + \mathcal O(\epsilon^4)
	\\
	\label{Zprime.Mass}
	m_{Z'}^2 & = \frac{1}{4} b^2 g_X^2 v_R^2 \left(1 + \epsilon^2\right) + \mathcal O(\epsilon^4),
\end{align}
where $\epsilon^2 = \frac{g_X^2 a^2 v^2}{g_X^2 b^2 v_R^2 - \left(g_1^2 + g_2^2\right) v^2} \ll 1$.
The most stringent bounds on the $Z'$ mass comes from LEP2 and depend on its
couplings to charged leptons. In this case then, it will depend on the value
of $a$ and $b$.  See Ref~\cite{Petriello:2008zr} for a recent study of
the $Z^{'}$ at the LHC and Ref~\cite{Langacker:2008yv} for a review.

The $Z$-$Z'$ mixing is also constrained and must be of order $10^{-3}$.  Its value can
be found by projecting out the  zero-mode photon from Eq.~(\ref{Z.Mass.Matrix})
and is given by:
\begin{equation}
	\tan{2 \xi} = 2 \frac{M_{Z Z'}^2}{M_{Z'}^2 - M_Z^2}
\end{equation}
where
\begin{align}
M_{Z Z'}^2 & = \frac{1}{4} \ a \ g_X \sqrt{g_1^2+ g_2^2} \ v^2
	\\
	M_Z^2 & = \frac{1}{4} \ (g_1^2 + g_2^2) \ v^2
	\\
	M_{Z'}^2 & = \frac{1}{4} \ a^2 \ g_X^2 \ v^2 + \frac{1}{4} \ b^2 g_X^2 \ v_R^2
\end{align}
Keeping up to first order in $\frac{v^2}{v_R^2} \sim \epsilon^2$ yields:
\begin{equation}
	\xi = \frac{a \sqrt{g_1^2 + g_2^2}}{b^2 \ g_X} \ \frac{v^2}{v_R^2},
\end{equation}
which can easily satisfy the bound.
{\subsubsection{\bf III.A.2. Neutralinos and Neutrinos}}
Once R-parity is broken the neutralinos and neutrinos mix.
Defining the basis
$\left(\nu, \ \nu^c, \ \tilde B^{'}, \ \tilde B,\ \tilde W_L^0, \ \tilde H_d^0, \ \tilde H_u^0 \right)$
their mass matrix is given by
\begin{widetext}
\begin{equation}
	{\cal M}_{N} =
	\begin{pmatrix}
			0
		&
			\frac{1}{\sqrt{2}} Y_\nu^D v_u
		&
			-\frac{1}{2} (a+b) g_{X} v_L
		&
			-\frac{1}{2} g_1 v_L
		&
			\frac{1}{2} g_2 v_L
		&
			0
		&
			\frac{1}{\sqrt{2}} Y_\nu^D v_R
	\\
			\frac{1}{\sqrt{2}} Y_\nu^D v_u
		&
			0
		&
			\frac{1}{2} b \ g_{X} v_R
		&
			0
		&
			0
		&
			0
		&
			\frac{1}{\sqrt{2}} Y_\nu^D v_L
	\\
			-\frac{1}{2} (a+b) g_{X} v_L
		&
			\frac{1}{2} b \ g_{X} v_R
		&
			M_{X}
		&
			0
		&
			0
		&
			-\frac{1}{2} a \ g_X v_d
		&
			\frac{1}{2} a \ g_X v_u
	\\
			-\frac{1}{2} g_1 v_L
		&
			0
		&
			0
		&
			M_1
		&
			0
		&
			-\frac{1}{2} g_1 v_d
		&
			\frac{1}{2} g_1 v_u
	\\
			\frac{1}{2} g_2 v_L
		&
			0
		&
			0
		&
			0
		&
			M_2
		&
			\frac{1}{2} g_2 v_d
		&
			-\frac{1}{2} g_2 v_u
	\\
			0
		&
			0
		&
			-\frac{1}{2} a \ g_X v_d
		&
			-\frac{1}{2} g_1 v_d
		&
			\frac{1}{2} g_2 v_d
		&
			0
		&
			-\mu
	\\
			\frac{1}{\sqrt{2}} Y_\nu^D v_R
		&
			\frac{1}{\sqrt{2}} Y_\nu^D v_L
		&
			\frac{1}{2} a \ g_X v_u
		&
			\frac{1}{2} g_1 v_u
		&
			-\frac{1}{2} g_2 v_u
		&
			-\mu
		&
			0
	\end{pmatrix}.
\label{neutralino}
\end{equation}
\end{widetext}
In order to understand the neutrino masses we focus on
the simple case $v_L \to 0$ and $Y_\nu^D$ small. Then,
in this limit the neutrino mass matrix is given by
\begin{equation}
M_\nu = M_\nu^I + M_\nu^R,
\end{equation}
where $M_\nu^I$ is the type I seesaw contribution~\cite{TypeI}
and $M_\nu^R$  is due to R-parity violation.
These contributions are given by
\begin{eqnarray}
	M_\nu^I	&=&	\frac{1}{2} Y_\nu^D M_{\nu^C}^{-1} \left(Y_\nu^D\right)^T v_u^2,
	\\
	M_\nu^R	&=&	m \ M_{\tilde \chi^0}^{-1} \ m^T,
\end{eqnarray}
where $m=diag \left( 0, 0, 0, 0, Y_\nu^D v_R / \sqrt{2}\right)$.
Therefore, it is possible to generate the neutrino masses in a consistent way.
Notice the possible strong mixing between $\tilde{B}^{'}$ and the Higgsinos.

\subsubsection{\bf III. A. 3. Higgses and Sleptons}
Defining the basis $\sqrt{2} \imag \left(\tilde \nu, \tilde \nu^c, H_d^0, H_u^0 \right)$
for CP-odd scalars, $\sqrt{2} \real \left(\tilde \nu, \tilde \nu^c, H_d^0, H_u^0 \right)$
for CP-even scalars and for the charged scalars $\left(\tilde e^*, \tilde e^c, H_d^{-*}, H_u^+ \right)$
the mass matrices are given by Eq.(\ref{CP-odd}), Eq.(\ref{CP-even}) and Eq.(\ref{Charged}),
respectively. The mass matrix for the CP-odd neutral Higgses reads as
\begin{widetext}
\begin{flushleft}
\begin{equation}
\label{CP-odd}
{\cal M}_{P}^2
	=
	\begin{pmatrix}
	  	\frac{v_R}{v_L} B_\nu
	&
		B_\nu
	&
		-\frac{1}{\sqrt{2}} Y_\nu^D \mu \ v_R
	&
		-\frac{1}{\sqrt{2}} A_\nu^D v_R
\\
		B_\nu
	&
		\frac{v_L}{v_R} B_\nu
	&
		-\frac{1}{\sqrt{2}} Y_\nu^D \mu \ v_L
	&
		-\frac{1}{\sqrt{2}} A_\nu^D v_L
\\
		-\frac{1}{\sqrt{2}} Y_\nu^D \mu \ v_R
	&
		-\frac{1}{\sqrt{2}} Y_\nu^D \mu \ v_L
	&
		\frac{v_u}{v_d} B\mu \ + \frac{Y_\nu^D \mu \ v_L v_R}{\sqrt{2} v_d}
	&
		B\mu
\\
		-\frac{1}{\sqrt{2}} A_\nu^D v_R
	&
		-\frac{1}{\sqrt{2}} A_\nu^D v_L
	&
		B\mu
	&
		\frac{v_d}{v_u} B\mu \ - \frac{A_\nu^D v_L v_R}{\sqrt{2} v_u}
	  \end{pmatrix},
\end{equation}
\end{flushleft}
\end{widetext}
while for the CP-even scalars one finds
\begin{equation}
\label{CP-even}
	{\cal M}_S^2	=
	\begin{pmatrix}
		S_{\nu}^2
		&
		S_{\nu H}^2
	\\
		\left(S_{\nu H}^{2}\right)^T
		&
		S_{H}^2
	\end{pmatrix},
\end{equation}
where $S_{\nu}^2$, $S_{\nu H}^2$ and $S_H^2$ are given in the appendix.
It is well-known that in the MSSM the tree level upper bound on the
lightest CP-even Higgs is $M_Z$ and one can satisfy the experimental
bound on the Higgs mass once the radiative corrections are included.

In the case of the charged Higgses the mass matrix is given by
\begin{equation}
\label{Charged}
	M_C^2 =
	\begin{pmatrix}
		C_e^2
		&
		C_{e H}^2
	\\
		\left(C_{e H}^2 \right)^T
		&
		C_{H}^2
	\end{pmatrix}.
\end{equation}
See the appendix for the definition of $C_{e}^2$, $C_{eH}^2$ and $C_{H}^2$. 
It is important to show that the spectrum of the theory is realistic and 
the expected Goldstone bosons exist. Here, we analyze the spectrum in the very illustrative
limit of zero mixing between the left- and right-handed sneutrinos,
\textit{i.e.} $Y_\nu^D, A_\nu^D \rightarrow 0$. In this limit,
Eq.~(\ref{MC.vR}) indicates that $v_L \rightarrow 0$ as well and
$B_\nu \rightarrow 0$, by definition. In this limit, the $\tilde \nu$
component of the CP-even mass matrix decouples as does the same component
in CP-odd mass matrix. This complex left-handed sneutrino then has the mass:
\begin{align}
	\label{mnuMass}
	\notag
	m_{\tilde \nu}^2 	= &
		M_{\tilde L}^2 - \frac{1}{8} \left(a+b\right) g_{X}^2 \left(b \ v_R^2 + a \ v_u^2 - a \ v_d^2\right)
	\\
		&- \frac{1}{8}\left(g_1^2 + g_2^2 \right) \left(v_u^2-v_d^2 \right).
\end{align}
The remaining three-by-three CP-even mass matrix has potentially large mixing between the right-handed
sneutrino and Higgses.  An approximate solution can be attained in the limit of large $\tan \beta$.
Here the heavier MSSM Higgs decouples leaving a mixing between the up-type Higgs and the
right-handed sneutrino. The resulting trace and determinant are identical to those of the
the neutral gauge bosons, Eq.~(\ref{Z.Mass.Matrix}), demonstrating that the lightest Higgs in this case,
as in the MSSM, is bounded by the $Z$ mass at tree-level. The mass of the mostly
right-handed sneutrino in this limit is that of the $Z'$, Eq.~(\ref{Zprime.Mass}).

Applying the zero sneutrino mixing limit to the CP-odd and charged Higgs sector shows that
those matrices decouple into three values: two eigenvalues representing the left- and right-handed
slepton masses and the MSSM two-by-two mass matrix for the up- and down-type Higgs. We will
focus on the former since the latter only reproduces the results of the MSSM.

Apart from the left-handed sneutrino mentioned above, the CP-odd matrix contains the Goldstone
boson eaten by $Z^{'}$ (the Majoron). It is completely made up of the imaginary part of
the right-handed sneutrino, $\imag \tilde \nu^c$. Finally, the masses of the charged sleptons are:
\begin{align}
\notag
	m_{\tilde e_L}^2 	= &
		M_{\tilde L}^2 - \frac{1}{8} \left(a+b\right)g_{X}^2 \left(b \ v_R^2 + a \ v_u^2 - a \ v_d^2\right)
\\
		& + \frac{1}{8} \left(g_2^2 - g_1^2 \right) \left(v_u^2 - v_d^2 \right) + \frac{1}{2} Y_e^2 v_d^2,
\\ \notag
	m_{\tilde e_R}^2 	= &
		M_{\tilde E^c}^2 + \frac{1}{8} \left(2a + b\right) g_{X}^2 \left(b \ v_R^2 + a \ v_u^2 - a v_d^2\right)
\\
		 & + \frac{1}{4} g_1^2 \left(v_u^2 - v_d^2 \right)+ \frac{1}{2} Y_e^2 v_d^2.
\end{align}
A closer examination of these approximate masses for the MSSM fields indicates that these values are the MSSM mass values modified appropriately by the $X$ $D$-term contributions. All of these masses are realistic given $M_{\tilde L}^2 
> \frac{1}{8} \left(a+b\right)g_{X}^2 \left(b \ v_R^2 + a \ v_u^2 - a \ v_d^2\right)$.
Of further interest is the prediction of the degeneracy between the $Z^{'}$ gauge boson and the
physical right-handed sneutrino. Corrections to the approximate masses presented here would be
 suppressed by neutrino masses, making this discussion relevant even in the non-limit case.
{\subsubsection{\bf III. A. 4. Charginos and Charged Leptons}}
Mixing between the charged leptons and the charginos will occur
in the charged fermion sector, $\left(e^c, \ \tilde W_L^+, \ \tilde H_u^+ \right)$ and
$\left( \ e, \ \tilde W_L^-, \ \tilde H_d^-\right)$. In this basis the mass matrix is
given by 
\begin{equation}
	{\cal M}_{\tilde C} =
	\begin{pmatrix}
			-\frac{1}{\sqrt{2}} Y_e v_d
		&
			0
		&
			\frac{1}{\sqrt{2}} Y_e v_L
	\\
			\frac{1}{\sqrt{2}} g_2 v_L
		&
			M_2
		&
			\frac{1}{\sqrt{2}} g_2 v_d
	\\
			-\frac{1}{\sqrt{2}} Y_\nu^D v_R
		&
			\frac{1}{\sqrt{2}} g_2 v_u
		&
			\mu
	\end{pmatrix}.
\end{equation}
Since the mixing between the MSSM charginos and the charged leptons is
proportional to $v_L$ and $Y_\nu^D$ small corrections to the charged lepton
masses can exist once the charginos are integrated out. However, this
contribution is always small once we impose the neutrino constraints.
\\
\section{IV. Collider Signals}

As a consequence of R-parity violation, the lightest neutralino will be
unstable and will decay via lepton number violating interactions. These type
of interactions will also exist for the charginos and the new gauge boson:

\textit{Sleptons decays and production at the LHC}: It is important to emphasize the lepton number violating decays of sleptons:
$\tilde{\nu} \ \to \nu \nu, e^+_i e^-_j$, $\tilde{e}_i \to e_j \nu_k$, $\tilde{e}^c_i \to e_j
\bar{\nu}^c_k$ and $\tilde{\nu^c} \to e \tilde{e}^c$. These decays
are proportional to $v_L$ or $Y_\nu^D v_R$ and are crucial for the test of the model.

The $Z'$ allows for a new production mechanism for sleptons at the LHC:
$$ pp \ \to \ Z, Z^{'} \ \to \  \tilde{\nu} \tilde{\nu}^*$$
Therefore, channels with four leptons in the final state are possible: $eeee$, $e \mu \mu \mu$,
$e e \mu \mu$, $e e e \mu$, $\mu \mu \mu \mu$ and also with several tau's. Then, one could test the existence
of R-parity violation and lepton number violation in this way. See Ref.~\cite{LTWang} for the
study of this production mechanism at the LHC.

\textit{$Z^{'}$ decays}: The $Z^{'}$ decays will be dependent on the values of $a$ and
$b$.  Determination of these values would be crucial to understanding the nature of the abelian
symmetry. In addition to the typical $Z^{'}$ decays,
new lepton number violating decays will be possible. These include
$Z^{'} \rightarrow e_j^\pm \tilde \chi_j^\mp$ which are suppressed by $v_L$.
Also possible are the very interesting decays $Z^{'} \to \overline{\nu^C} \nu^C$,
where the right-handed neutrinos can decay mainly to an electron and a selectron.
These decays are lepton number violating and proportional to $v_R$.

\textit{Neutralino decays}: As it is well-known in the case
of R-parity violation the lightest neutralino is unstable.
These fields will decay as $\tilde \chi_i^0 \rightarrow Z^0 \bar \nu$ and
$\tilde \chi_i^0 \rightarrow W^\pm e^\mp$. In the case when
the neutralino is the up-like Higgsino, these decays are proportional
to $v_R$, while in the rest of the cases are suppressed by $v_L$.
For a recent study of these decays in SUSY models with R-parity violation 
see for example~\cite{Marco}.

\textit{L-violating Higgs decays}: The Higgses now has lepton number
violating decay channels open such as MSSM-like Higgs into a slepton and
a $W$ or $Z$ if kinematically allowed.  More interesting is its mixing
with the right-handed sneutrino. This could change the Higgs decay 
into two gammas branching ration over its decay to matter branching ratio.

\textit{Chargino decays}: In this case new decays into charged
leptons and a $Z$ or $W$ exist. In this case all these decays
are suppressed by $v_L$ or $Y_\nu^D$ once we impose
the constraints coming from neutrino masses.

It is important to emphasize that in order to test this model
at the LHC one should discover the $Z^{'}$, the right-handed
neutrinos crucial to cancel anomalies, and understand the lepton 
number and R-parity violating decays of the sneutrinos. 
\section{V. Other Aspects}
As it is well-known in this context the gravitino 
can be a dark matter candidate\footnote{
It is a common misbelief that once R-parity in broken
in supersymmetric theories one does not have
a cold dark matter candidate. Fortunately,
it is not always the case since if the
gravitino is the lightest supersymmetric particle it
still can be a good cold dark matter candidate
since its decay rate will be suppressed by the Planck
scale and the R-parity violating couplings. A naive estimation
gives us
\begin{equation}
\tau_{3/2} \sim 10^{26} \ \text{s} \ \left(\frac{\epsilon}{10^{-7}}\right)^{-2} \ \left( \frac{m_{3/2}}{10 \ \text{GeV}}\right)^{-3},
\end{equation}
where $m_{3/2}$ is the gravitino mass. $\epsilon$ defines the amount of R-parity violation, 
and it is proportional to the ratio between the R-parity violating coupling and the soft mass.
See Ref.~\cite{Yanagida} for the study of gravitino dark matter in R-parity violating scenarios.}.
Here we discuss the issue of proton decay.

\subsection{V.II. Proton Stability}
Let us discuss the possible constraints coming
from proton decay. See Ref.~\cite{Nath} for a
review on proton decay. There are several
operators which are relevant for proton decay
in this context. One has the so-called LLLL
dimension five contributions
\begin{equation}
{\cal O}_{LLLL}= \alpha_{ijkl} \ \hat{Q}_i \ \hat{Q}_j \ \hat{Q}_k \ \hat{L}_l / \Lambda_B,
\end{equation}
and the RRRR contributions
\begin{equation}
{\cal O}_{RRRR}= \beta_{ijkl} \ \hat{U}^C_i \ \hat{D}^C_j \ \hat{U}^C_k \ \hat{E}^C_l / \Lambda_B.
\end{equation}
See Ref.~\cite{Proton-SUSY} for the possibilities to suppress these contributions.
Now, in this context one has an extra operator due to the existence
of the right-handed neutrinos
\begin{equation}
{\cal O}_{RRRR}^\nu = \gamma_{ijkl} \ \hat{U}^C_i \ \hat{D}^C_j \ \hat{D}^C_k \ \hat{N}^C_l / \Lambda_B.
\end{equation}
Here $\gamma_{ijkl}=-\gamma_{ikjl}$. In this case when the ``right-handed" sneutrino gets a
VEV, baryon number violating interactions present in the MSSM are generated. Then, one finds
$u^C_i \ d^C_j \ \tilde{d}^C_k \ v_{Rl} / \Lambda_B$, $u^C_i \ \tilde{d}_j^C \ d_k^C \ v_{Rl}/ \Lambda_B$,
and $\tilde{u}^C_i \ d_j^C \ d_k^C \ v_{Rl}/ \Lambda_B$. Using the new interactions in the
superpotential proportional to $Y_\nu^D$ and the above operator one finds the following
constraint coming from proton decay (as example we use $p \to \pi^0 e^+$):
\begin{equation}
\gamma_{112i} \frac{(Y_\nu^D)^{1j} \ Y_s}{\Lambda_B} \frac{v_R^i v_R^j}
{M_{\tilde{\chi}^o} \ M_{\tilde{s}^C}^2} \ \lesssim \ 10^{-30} \ \text{GeV}^{-2}.
\end{equation}
Then, it is easy to show that if $\Lambda_B \sim M_{Pl}$ the coupling $\gamma_{112i}$ can be
of order one. See Ref.~\cite{FileviezPerez:2004th} for the constraints coming from different channels.
\section{VI. Summary and Outlook}
We studied a consistent and general mechanism for spontaneous R-parity
violation in a class of simple extensions of the minimal supersymmetric standard
model (MSSM) with an extra Abelian gauge symmetry which is a linear
combination of $B-L$ and weak hypercharge. In this case we found
that this mechanism can be realized even with positive soft masses
for ``right-handed" sneutrinos, whose VEV breaks both $U(1)_{X}$ and R-parity. A realistic
mechanism for generating neutrino masses exists as well as a realistic spectrum.
We briefly discussed the possible collider signals which could be
used to test the theory, contributions for proton decay
and the gravitino as a dark matter candidate.

{\textit{Acknowledgments}}:
The work of P.F.P. was supported in part by the U.S.
Department of Energy contract No. DE-FG02-08ER41531 and in part
by the Wisconsin Alumni Research Foundation. S.S. is supported in
part by the U.S. Department of Energy under grant No. DE-FG02-95ER40896,
and by the Wisconsin Alumni Research Foundation.
\appendix
\section{APPENDIX: Mass Matrices}
Using the basis $\sqrt{2} \real \left(\tilde \nu, \tilde \nu^c, H_d^0, H_u^0 \right)$
for CP-even scalars and for the charged scalars $\left(\tilde e^*, \tilde e^c, H_d^{-*}, H_u^+ \right)$,
the mass matrices for these two sectors are:
\begin{equation}
	{\cal M}_S^2	=
	\begin{pmatrix}
		S_{\nu}^2
		&
		S_{\nu H}^2
	\\
		\left(S_{\nu H}^{2}\right)^T
		&
		S_{H}^2
	\end{pmatrix},
\end{equation}
and
\begin{equation}
	M_C^2 =
	\begin{pmatrix}
		C_e^2
		&
		C_{e H}^2
	\\
		\left(C_{e H}^2 \right)^T
		&
		C_{H}^2
	\end{pmatrix},
\end{equation}
where:
\begin{widetext}
\begin{align}
	S_\nu^2	\equiv	&
	\begin{pmatrix}
		\frac{1}{4} \left(g_1^2 + g_2^2 + \left(a+b\right)^2 g_{X}^2 \right) v_L^2 + \frac{v_R}{v_L} B_\nu
		&
		-\frac{1}{4} \left(\left(a+b\right)b \ g_{X}^2 - 4 Y_\nu^{D2}\right) v_L v_R - B_\nu
	\\
		-\frac{1}{4} \left(\left(a+b\right)b \ g_{X}^2 - 4 Y_\nu^{D2}\right) v_L v_R - B_\nu
		&
		\frac{1}{4} b^2 g_{X}^2 v_R^2 + \frac{v_L}{v_R} B_\nu
	\end{pmatrix},
\\
\nonumber
\\
	S_{\nu H}^2 \equiv	&
	\begin{pmatrix}
		\frac{1}{4} \left(g_1^2 + g_2^2 + a \left(a+b \right) g_X^2\right) v_d v_L  - \frac{1}{\sqrt{2}} Y_\nu^D \mu v_R
		&
		-\frac{1}{4} \left(g_1^2 + g_2^2  + a \left(a+b\right) g_X^2  - 4 Y_\nu^{D2} \right) v_L v_u + \frac{1}{\sqrt{2}} A_\nu^D v_R
	\\
		-\frac{1}{\sqrt{2}} Y_\nu^D \mu v_L - \frac{1}{4} a \ b \ g_X^2 v_d v_R
		&
		\frac{1}{4} a \ b \ g_X^2 v_u v_R + {Y_\nu^D}^2 v_u v_R + \frac{1}{\sqrt{2}} A_\nu v_L
	\end{pmatrix},
\\
\nonumber
\\
	S_{H}^2	\equiv &
	\begin{pmatrix}
		\frac{1}{4} \left( g_1^2 + g_2^2 + a^2 g_X^2\right)v_d^2 + \frac{v_u}{v_d} B\mu  + \frac{Y_\nu^D \mu v_L v_R}{\sqrt{2} v_d}
		&
		 - \frac{1}{4} \left(g_1^2 + g_2^2 + a^2 g_X^2\right) v_u v_d - B\mu
	\\
		- \frac{1}{4} \left(g_1^2 + g_2^2  + a^2 g_X^2\right) v_u v_d - B\mu
		&
		\frac{1}{4} \left(g_1^2 + g_2^2  + a^2 g_X^2\right) v_u^2 + \frac{v_d}{v_u} B\mu - \frac{A_\nu^D v_L v_R}{\sqrt{2} v_u}
	\end{pmatrix}.	
\end{align}

\begin{flushleft}
\begin{align}
	C_e^2 \equiv &
	\begin{pmatrix}
		C_{11}^2
		&
		B_e
	\\
		B_e
		&
		C_{22}^2
	\end{pmatrix},
\\
\nonumber
\\
	C_{e H}^2 \equiv
	&
	\begin{pmatrix}
		\frac{1}{4} g_2^2 v_d v_L - \frac{1}{2} Y_e^2  v_d v_L - \frac{1}{\sqrt{2}} Y_\nu^D \mu v_R
		&
		\frac{1}{4} g_2^2 v_L v_u - \frac{1}{2} Y_\nu^2 v_u v_L - \frac{1}{\sqrt{2}} A_\nu^D v_R
	\\
		\frac{1}{2} Y_e Y_\nu^D v_u v_R + \frac{1}{\sqrt{2}} A_e v_L
		&
		\frac{1}{2} Y_e Y_\nu^D v_d v_R + \frac{1}{\sqrt{2}} Y_e \mu v_L
	\end{pmatrix},
\\
\nonumber
\\
	C_{H}^2 \equiv
	&
	\begin{pmatrix}
		\frac{1}{4} g_2^2\left(v_u^2-v_L^2 \right) + B\mu \frac{v_u}{v_d} + \frac{1}{2} Y_e^2 v_L^2 + \frac{Y_\nu^D \mu v_R v_L}{\sqrt{2} v_d}
		&
		B\mu + \frac{1}{4} g_2^2 v_u v_d
	\\
		B\mu + \frac{1}{4} g_2^2 v_u v_d
		&
		\frac{1}{4} g_2^2 \left(v_d^2 + v_L^2 \right) + \frac{v_d}{v_u}B\mu - \frac{1}{2} Y_\nu^{D2} v_L^2 - \frac{A_\nu^D v_L v_R}{\sqrt{2} v_u}
	\end{pmatrix}.
\end{align}
\end{flushleft}

In the above equations $C_{11}^2$ and $C_{22}^2$ are given by
\begin{eqnarray}
C_{11}^2 & = & \frac{1}{4} g_2^2 \left(v_u^2 - v_d^2 \right) + \frac{1}{2}Y_e^2 v_d^2 - \frac{1}{2} {Y_\nu^D}^2 v_u^2 + \frac{v_R}{v_L} B_\nu,
\end{eqnarray}
and
\begin{eqnarray}
C_{22}^2 &=& M_{\tilde E^c}^2 + \frac{1}{4} g_1^2 \left(v_u^2 - v_d^2 -v_L^2 \right)
			+ \frac{1}{8} \left(2a+b\right)g_{X}^2 \left(b \ v_R^2 + a \ v_u^2 - a \ v_d^2 - \left(a+b\right)v_L^2 \right)
			+ \frac{1}{2} Y_e^2 \left( v_d^2 + v_L^2 \right).
\end{eqnarray}
We also define for convenience:
$$
	B_\nu = \frac{1}{\sqrt{2}} \left(Y_\nu^D \mu \ v_d - A_\nu^D v_u \right),
	\text{ and }
	B_e = \frac{1}{\sqrt{2}} \left(Y_e \ \mu \ v_u - A_e v_d \right).
$$
\end{widetext}


\end{document}